\documentclass[sigconf,nonacm,authorversion]{acmart}

\usepackage{amsmath,amsfonts}
\usepackage{algorithmic}
\usepackage{graphicx}
\usepackage{textcomp}
\usepackage{xcolor}
\usepackage{soulpos}
\usepackage{graphicx}
\usepackage{multirow}
\usepackage{listings}
\usepackage{subcaption}
\usepackage{minted}
\usepackage{ragged2e}
\usepackage{multirow}
\usepackage{color,soul}
\def\BibTeX{{\rm B\kern-.05em{\sc i\kern-.025em b}\kern-.08em
    T\kern-.1667em\lower.7ex\hbox{E}\kern-.125emX}}

\begin{document}

\title{Scaling Up Throughput-oriented LLM Inference Applications \\on
Heterogeneous Opportunistic GPU Clusters \\with Pervasive Context Management}

\author{Thanh Son Phung, Douglas Thain}
\affiliation{
    \institution{University of Notre Dame}
    \city{Notre Dame}
    \country{USA}}
\email{{tphung, dthain}@nd.edu}

\settopmatter{printacmref=false}

\begin{abstract}

The widespread growth in LLM developments increasingly demands more computational power from clusters than what they can supply. Traditional LLM applications inherently require huge static resource allocations, which force users to either wait in a long job queue and accept progress delay, or buy expensive hardware to fulfill their needs and exacerbate the demand-supply problem.
However, not all LLM applications are latency-sensitive and can instead be executed in a throughput-oriented way. This throughput orientation allows a dynamic allocation that opportunistically pools available resources over time, avoiding both the long queue and expensive GPU purchases. Effectively utilizing opportunistic resources brings numerous challenges nevertheless. Our solution, pervasive context management, exploits the common computational context in LLM applications and provides mechanisms and policies that allow seamless context reuse on opportunistic resources. Our evaluation shows an LLM application with pervasive context management on opportunistic resources reduces its execution time by 98.1\%.

\end{abstract}

\maketitle

\keywords{Large Language Models; Machine Learning Inferences; Workflow Systems; Serverless Computing; Burst Buffers}

\section{Introduction}
\label{sec:intro}

AI is expected to become the next revolution in human productivity, attracting both significant interests and huge investments from governments, industries, and the academia around the world\cite{fasst,invest-300b-ai,2023-taulbee-survey}.
Most of these interests and investments concentrate on Generative AI (GenAI), a type of AI that creates new data based on patterns implicitly extracted from a huge amount of existing data. 
Large Language Models (LLMs) are the current powerhouse of GenAI, boasting astonishing generative capabilities in diverse subjects (e.g., language, mathematics, coding) and various formats (e.g., text, image, video), on par with or even surpassing the average human intelligence\cite{team2023gemini, achiam2023gpt, claude3}. However, LLMs need a massive amount of computational power to tune its parameters (i.e., train) and create new data (i.e., infer), which can only be offered at the moment by 1) commercial general-purpose GPUs\cite{amdgpu, nvidiagpu}, which are in high demand, or 2) custom ASIC or FPGA accelerators\cite{jouppi2017datacenter, firoozshahian2023mtia}, which are often proprietary and extremely complex to design, both of which are limited and expensive. Thus, while GenAI researchers and practitioners are optimistic about the infinite possibilities of LLM applications, system administrators are instead cautious in promising SLOs/SLAs due to the finite amount of computational resources, pressing the need for efficient resource management.

To realize a solution for this need, one needs to be conscious of different resource requirements that different modes of LLM developments require.
LLM computations revolve around either training or inference. LLM training typically requires a static allocation of resources in a cluster for a long period of time, which  is usually satisfied by a dedicated cluster or a set of machines carved out of a shared pool of resources. This siloing approach ensures that the training process is uninterruptible and minimizes the turnaround time, but introduces an unavoidable waste from 
idle resources during the process\cite{gao2024empirical, delestrac2024multi}.  
Inference serving is latency-sensitive as it directly interacts with end users, which also requires a static allocation of resources in an indefinite amount of time and  suffers from expenses caused by the idle resource problem\cite{wei2022gpu, jin2023s}. Remedies\cite{jiang2024megascale,hisaharo2024optimizing,hayashi2024enhancing,wang2023tabi} have been proposed and implemented to alleviate this problem, but the root cause of transient idle resources can never be eliminated due to the nature of resource silos and the priority and urgency of LLM training and inference serving. 
Moreover, the high priority of  LLM jobs can worsen the oversubscription of jobs in a cluster, creating a long queue time on the scale of hours, days, or even weeks. This inadvertently pushes users with tough deadlines away from sharing resources to buying their exclusive hardware, further exacerbating the market price of GPUs and the idle resource problem.

We observe that many (LLM) inference applications are not inherently latency-sensitive but can instead be more throughput-oriented
(e.g., model research and validation, prompt engineering, analytics).
A throughput-oriented model of execution allows a dynamic allocation of resources in which the current load of the cluster dictates the allocation or preemption of resources:  resource allocation gradually climbs as more resources become available and gradually drops to make room for a more prioritized job; the application throughput should simply adjust proportionally. This resource dynamicity thus puts throughput-oriented inference applications out of the long static-allocation queue and allows the opportunistic harvesting of unallocated resources as a result of the cluster's resource fragmentation. Therefore, we believe the key to scale up throughput-oriented inference applications is to run them in an opportunistic way.

Of course, opportunistic resource utilization comes with a unique set of problems. Such resources are unstable: resources can be evicted at any moment, threatening ongoing progress to be uselessly discarded, and resources can come at any moment, requiring an application to adjust quickly to the newly available resources. 
These resources are  unpredictable: they can vary day-by-day, or even hour-by-hour, and only correlate with the current load of the cluster. 
Opportunistic resource acquisition ensures runtime execution as long as they are held, but that doesn't guarantee throughput: each computational unit usually brings non-trivial initialization and cleanup overheads, and these overheads, if not carefully managed, can substantially degrade the application's performance and the overall cluster's efficiency as computations may repeatedly run   without any throughput.

Resource heterogeneity is often common in opportunistic resources: a typical cluster has a specific spectrum of slow and fast GPUs as it evolves over a long period of time, and while a user can ask for a static allocation of homogeneous GPUs, one cannot simply pick and choose opportunistically available GPUs as they may come and go in arbitrary orders and varieties. 
Data movement and I/O can also be spiky  as computational units often start and stop arbitrarily in response to the instability of opportunistic resources. Finally,  it is unclear how a user can choose a batch size for each computational unit: a batch size too large unlocks a higher inference throughput but risks a higher chance of eviction and thus no throughput, and a batch size too small safeguards the incremental smaller throughput but wastes resources from the accompanying initialization and cleanup overheads.

Our insight into unlocking the benefits of opportunistic resource utilization and negating its drawbacks is as follows: {\bf individual computational units that batch multiple inferences often share the same computational context (e.g., software dependencies, model weights,  prompt template), and managing the reuse of this context effectively is the key to productive opportunistic resource utilization.} Such context management must be pervasive: a computational context can reside in memory, local SSDs, or GPUs of any compute node in the allocated pool of opportunistic resources. Once this context is captured and registered to be managed by a given system, it is then straightforward to derive policies and mechanisms that adjust in real-time the distributions, 
retentions, and reuses of contexts   to the organic arrival and departure of opportunistic resources. Our evaluation shows that, compared to the baseline of running a given inference application on a dedicated high-quality GPU, a user can reduce the execution time of that application by 98.1\% (from 11.4 hours to 13.1 minutes)  by carefully and effectively harvesting opportunistic resources in a GPU cluster. Our evaluation also shows that an inattentive solution trying to utilize opportunistic resources incorrectly leads to a terrible degradation of performance and increases the execution time of the inference application by 245.3\% (from 11.4 hours to 1.6 days).

\section{Background}
\label{sec:background}

\subsection{Throughput-oriented Applications}
\label{subsec:xput-apps}
Throughput-oriented applications have been one of the most well-studied subjects in the traditional HPC literature\cite{tovar2017job, phung2024adaptive, sly2024reshaping, montero2011elasticity, teodoro2013high}. Almost all resource managers\cite{thain2005distributed, yoo2003slurm, k8s, gentzsch2001sun} support this mode of execution via the concept of job priority and mechanisms like job eviction and  pausing, and many runtime systems\cite{sly2023taskvine, babuji2019parsl, turilli2018building, deelman2015pegasus} have been developed to help formulate and scale up applications in this paradigm. While their usage covers a wide range of fields, these applications have a limited set of common denominators as follows:

\textbf{Bulk data processing demand}. Thoughput-oriented applications typically tackle data-intensive problems, both in size (on the order of TBs or PBs) and in number (e.g., many small data items).

\textbf{Divisible and parallelizable computation}. Such problems are addressed by dividing the whole computational need into many smaller units, each of which can fit on a compute node, and parallelizing on multiple compute nodes in a given resource allocation.

\textbf{Per-task latency tolerance}.
Users running throughput-oriented applications don't concern with the latencies of individual computational units (i.e., tasks), but instead on the application (i.e., job) as a whole. Thus, minor deviations in tasks' latencies are tolerated.

\textbf{Job delay tolerance}. Users generally don't have a hard deadline for jobs, but instead worry about the throughput rate: as long as the rate is within an acceptable range, users are satisfied.

\textbf{Inter-task independence}. Each task processes its own chunk of data and has no synchronization with other tasks. Thus, tasks are independent from one another: a failing task doesn't crash others. 

\textbf{Per-task fault tolerance}. 
Common policies encoded in many resource managers allow them to preempt throughput-oriented jobs for a higher-priority ones. Therefore, tasks are designed to take this practice into account and tolerate external faults.

\textbf{High batch factor}. Tasks are regularly batched with many computations to amortize the non-trivial initialization and cleanup overheads and increase their overall throughputs.

\vspace{-.1in}

\subsection{Current LLM Training and Inference Serving Approaches}
\label{subsec:llm-training-infer}
\subsubsection{LLM Training}
Currently, there are three main approaches to parallelism in LLM training: data parallelism, pipeline parallelism, and tensor parallelism. 
Note that these approaches commonly work in tandem with each other to parallelize the training process.

\textbf{Data Parallelism}. This approach\cite{jiang2024megascale} parallelizes the training process by distributing different chunks of input data to different GPUs for gradient computation. Specifically,
each GPU in a resource allocation gets its own copy of a given model's parameters and optimizer states. Each GPU is then given a different input batch, makes a 1-forward-1-backward pass on the model with the batch to compute a local gradient, exchanges its gradient with all other GPUs in a full-mesh manner, and updates its model's parameters. Of course, a model might be too big for a GPU's memory, necessitating latter approaches that slice the model parameters in different ways.

\textbf{Pipeline Parallelism}. This approach\cite{NEURIPS2019_093f65e0} "horizontally" slices a given LLM into multiple partitions that correspond to the layers in the LLM architecture, and hosts each partition on a GPU. A new input batch is processed sequentially by the partition order in both forward and backward passes, where the previous GPU sends relevant information to the next GPU in line. Since this sequencing is blocking - only 1 GPU is active at a time - an input batch is instead split into many smaller mini-batches, creating a pipeline between GPUs. Upon receiving gradients of all mini-batches, the first GPU in line broadcasts them to all other GPUs and updates its model.

\textbf{Tensor Parallelism}.
This approach\cite{shoeybi2019megatron} "vertically" slices a given LLM's layer into multiple partitions, each of which is hosted on a GPU. An input data is also sliced accordingly, where each portion of the input is run in parallel through the associated parameter partition on a GPU, and the results of a layer between GPUs are synchronized at the end of the layer.
\vspace{-.05in}

\subsubsection{Inference Serving}
LLM inference serving resembles LLM training as it only makes a forward pass of the model to generate a prediction, and thus shares the pipeline and tensor parallelism approaches\cite{li2024llm}. A distinct 
characteristic of LLM inference serving is its use of the KV cache to store representations of the previously generated tokens and allow a linear growth of computation per output token. Without the KV cache, an LLM must repeatedly tokenize all input texts per output token which results 
in a quadratic growth of computation with the length of the output text.

The use of the KV cache of LLM inference serving draws two observations. First, the number of cached tokens (i.e., the token context window) can be too large to fit in a GPU. For example, two recently deployed LLM models, Gemini 2.5 \cite{gemini25} and Llama 4 Scout\cite{llama4scout}, have respective token context windows of 1 million and 10 million. Therefore, this requires a distributed approach to store the cache in multiple GPUs, which involves cross-GPU communication and synchronizations. 
Second, the LLM inferencing is commonly broken down into 2 different phases: a pre-fill phase and a decode phase. This separation comes from the fact that the cost of producing the first token is significantly higher than other output tokens as the whole prompt must be tokenized and cached in the process. Therefore, these two phases are usually executed on 2 different set of GPUs, where one set pre-fills the KV cache and then sends it to the other set of GPUs to continue token generations.

\subsubsection{Resource Requirements}
As described above, LLM training and inference serving exhibit a high degree of synchronization and cross-GPU communications. LLM serving has an even stronger latency requirement as it directly faces end-users, requiring 2 different sets of GPUs where one bootstraps the KV cache to be used by the other. Resource-wise, such tight coupling mode of computation mandates LLM training and inference serving to run on a large static allocation, and node failure is the exception rather than the norm. 
Techniques like checkpointing help preserve the work done but introduce high I/O and additional synchronization overheads and waste even more idle resources in the process. 

However, the impacts made by LLM innovations are irrefutable as mentioned in Section \ref{sec:intro}. LLM training and inference serving jobs are hence placed in the highest priority in clusters, each of which now has a few but huge jobs occupying the majority of resources. Other users wishing to use multiple GPUs in their jobs either have to wait for days or weeks until these jobs finish and quickly fill in the gap, or buy more hardware to fulfill their needs. Both options are undesirable: the former is cheaper but blocks progress, and the latter allows progress but is much more expensive.

\section{Throughput-oriented Inference Applications}
\label{sec:xput}
The LLM computing landscape turns out to be quite extreme as users with lower-priority projects are stuck with two rather disheartening options. 
This section provides the rationale and three conditions for a third option that neither requires users to wait for their turn in a limited GPU cluster nor incentivizes users to purchase their own hardware.
We first identify a class of LLM inference applications that has a less stringent resource footprint than traditional large-scale LLM training and inference serving, and provides a way to seamlessly pack these applications into an already busy GPU cluster.

\subsection{Inferencing as Throughput-oriented Applications}
\subsubsection{Advances in Smaller LLMs}
Growth in LLM developments is mostly driven by the promise of unlimited use cases and supported with evidence of incremental improvements on various AI benchmarks\cite{phan2025humanity, white2024livebench, chiang2024chatbot}.
However, over the last few years, LLMs have reached a critical parameter mass of O(1 trillion) due to the computational bottleneck, and many research and initiatives have been undertaken to reduce the model size\cite{zhang2024tinyllama, javaheripi2023phi, fu2023specializing}. 
Besides financial benefits from the reduction, these works typically trade the generality of bigger LLMs for better performance and usability in specific use cases (e.g., domain specialization, model distillation, LLM interpretability). This push for model reduction also increases the accessibility to the LLM technology: instead of requiring access to tens of thousands of GPUs and a customized software stack backed by divisions of engineers, a standalone user with only a handful number of GPUs can build upon this technology and create new applications, which in turn increases the overall rate of innovation in the field. Thus, we arrive at the first condition: 

\textbf{Condition \#1:} An inference application uses a family of smaller LLMs as the backbone. Such LLMs have at most a dozen billion parameters that can fit nicely in a small number of GPUs.

\subsubsection{Throughput-oriented Computational Need}
Not all inference applications have to be interactive, especially those that use LLMs as an automation tool in a data processing pipeline. This relaxed latency requirement, combined with condition \#1 above, allows many inference applications to neatly distribute their delay-tolerant work on individual compute nodes and thus exhibit seven throughput-oriented characteristics as described in Section \ref{sec:background}. 
This class of inference applications can thus be executed on a dynamic allocation of resources that shrinks or expands according to the current load of a given cluster. Moreover, this resource dynamicity feature  uniquely allows throughput-oriented inference applications to receive resources from not only opportunistic allocations that come without any promise, but also any existing associated static allocation. 
Thus, we arrive at the second condition:

\textbf{Condition \#2:} An inference application can satisfy a given computational need in a throughput-oriented way.

\subsection{Opportunistic Resource Utilization}
Performance studies on clusters\cite{8425189, 8891022, 10.1145/2391229.2391236, 6253523} show that the average resource utilization is never close to 100\%, leaving many resources stranded at any given time. There are two main sources of stranded resources: external fragmentation and internal fragmentation. 
External fragmentation occurs in a cluster when a set of statically allocated jobs does not perfectly fit the cluster's resource capacity, resulting in unallocated resources. Internal fragmentation occurs when a job does not fully utilize all resources in its allocation, resulting in idle resources. Resource managers are fully aware of both types of resource inefficiency and commonly mitigate them via two solutions: backfilling and overcommitment. While backfilling alleviates the unallocated resources by scheduling small lower-priority jobs to fill the void, overcommitment allows a cluster to pack more jobs than it usually can, hoping that jobs don't consume all of their allocations at all times.
This common awareness is the third condition: well-maintained clusters, with varying degrees, will always have backfilling and overcommitment enabled, thus opening the door of opportunistic resource utilization to any jobs that are designed to consume them. We arrive at the final condition:

\textbf{Condition \#3}: Clusters rarely reach 100\% resource utilization, thus providing jobs with opportunistic resources from unallocated or unused allocation.

Once these three conditions are upheld, a throughput-oriented inference application using a family of smaller LLMs can freely utilize available opportunistic resources without waiting for its turn in the static allocation queue or purchasing expensive hardware.

\section{Scaling Challenges on Opportunistic Resources}
\label{sec:challenges}
Utilizing opportunistic resources brings a unique set of challenges however as they come without any guarantee. This section dissects the characteristics of opportunistic resources, challenges for applications to utilize them, and potential approaches to these challenges.

\textbf{Challenge \#1: Instability.} Opportunistic resources are inherently unstable as they can join and leave the resource pool at any given moment. Resource managers commonly reclaim them by evicting a running job on a compute node, and tasks running on this node are killed without any chance of cleanup. 
Mitigations for this problem include designing fault-tolerance tasks such that a task crashing mid-execution does not affect the overall application and can later be rescheduled on other nodes for execution. On the other hand, an application should respond quickly to the availability of new resources as those resources are at best claimed by other jobs and at worst idle in the process.
Thus, an application needs to run a daemon-like process that periodically monitors the availability of the cluster and reactively submits more small backfilling jobs.

\textbf{Challenge \#2: Unpredictability.} Availability of opportunistic resources is generally unpredictable as it correlates with the current load of the cluster at a given time. The larger the cluster, the more unpredictable it becomes, so job planning for this resource is virtually unreliable. This can only be alleviated by observability tools that transparently inform users of the current rate of throughput and the overall progress of the application.

\textbf{Challenge \#3: No throughput guarantee.} Acquiring an opportunistic resource allocation ensures that the application and its constituent tasks are running, but it does not guarantee any throughput. This is because tasks, especially LLM inferencing, have a non-trivial initialization overhead: an LLM's parameters must be staged to  a compute node's SSDs and/or memory before being loaded into a GPU. This overhead, combined with individual tasks' runtime, risks tasks' preemption before they deliver any goodput. Mitigations include amortization of this cost by locally caching models on compute nodes and tuning the inference batch size based on a specific eviction risk model.

\textbf{Challenge \#4: Resource heterogeneity.}
Clusters are heterogeneous as they evolve over a long period of time, thus consist of a specific mixture of slow and fast GPUs. Applications cannot ask for an allocation of fast GPUs as opportunistic resources come and go with arbitrary orders and varieties, and users instead must be aware of this inherent heterogeneity when designing applications.

\textbf{Challenge \#5: Spiky data movement and I/O.}
It is common for applications to rely on the existence of a shared filesystem in a cluster to stage input data to and output data from compute nodes.
Challenge \#1 implies that this reliance will introduce a much more erratic data movement and I/O patterns as tasks start and stop based on the cluster's availability. 
A possible scenario is when a large amount of resources suddenly become available, and tasks deployed on this opportunistic allocation all try to read billions of parameters out of the shared filesystem at once, which deteriorates the shared filesystem's overall health
and hurts co-located users. Resolving this problems requires an effective management of the distribution and caching of data on compute nodes.

\textbf{Challenge \#6: Unclear optimal batch size.}
Finally, the inherent uncertainty in opportunistic resource
utilization 

makes the problem of choosing an inference batch size to maximize throughput even more difficult. It's also not clear if a user should use one batch size for all tasks when factored in resource heterogeneity and erratic data movement. A trial-and-error approach can be used to mitigate this problem and gradually narrow down the range of an optimal batch size, but Challenge \#2 complicates this
due to the frequent state change of opportunistic resources.

\vspace{-.07in}

\section{Pervasive Context Management}
\label{sec:solution}
This section presents our solution, pervasive context management, to the problem of scaling up throughput-oriented LLM inference applications on heterogeneous opportunistic GPU clusters. Our solution comprises of three components: 1) a software stack of dynamic workflow systems that allows users to express their computational needs via intuitive  abstractions and addresses Challenges \#1 and \#2, (2) the pervasive context management technique that allows efficient reuse of computational context between tasks and addresses Challenges \#3, \#5, and \#6, and (3) supporting mechanisms and policies that provide an end-to-end performant execution of these applications and address Challenge \#4.

\subsection{Parsl-TaskVine Software Stack}

The first component of our solution is the software stack of two dynamic workflow systems - Parsl\cite{babuji2019parsl} and TaskVine\cite{sly2023taskvine}. Parsl is a Python-native parallel library that allows users to express their computational needs via generic Python functions and automatically scales the computation on thousands of compute nodes. It excels in flexibility, portability, and ease of use, and is the default runtime of the Globus Compute ecosystem\cite{bauer2024globus}. 
TaskVine is a low-level data-intensive workflow execution engine. 
TaskVine's main strengths are its rich APIs that allow users to express low-level details about tasks and their inter-relationships, intelligent scheduling and optimization algorithms that extract values from these details, and novel data-intensive capabilities that cater to large-scale data processing applications. These strengths therefore considerably accelerate such applications to complete their executions in a near real-time manner\cite{sly2024reshaping, phung2024accelerating}. The software stack thus reflects our work on combining the best of both systems without sacrificing ease of use or performance.

\begin{figure}[t]
    \includegraphics[width=\columnwidth]{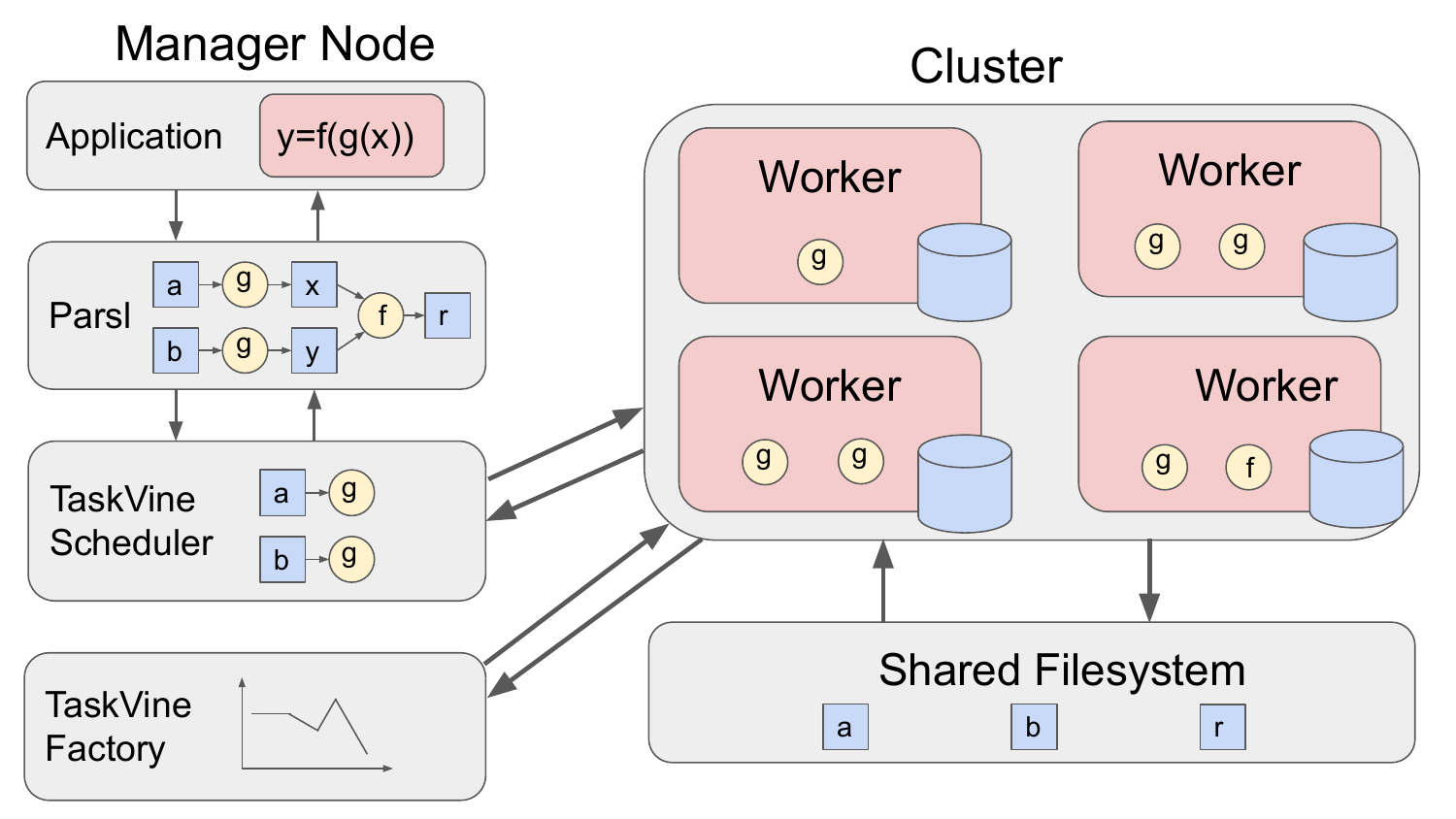}
    \caption{Software Stack of Dynamic Workflow Systems}
    \label{fig:software-stack}
    \vspace{-.18in}
\end{figure}

Figure \ref{fig:software-stack} shows how these two workflow systems work together in the big picture.
On the manager node, a user expresses their application's computational needs via generic Python functions. Once the application is run and these functions are invoked, they are intercepted and passed to Parsl for inter-function dependency management and function-to-task translation. Parsl sends ready tasks to the TaskVine scheduler, where they are examined for common execution and I/O patterns and scheduled for execution on workers accordingly. 
The TaskVine scheduler manages resources in the system via TaskVine workers, where each worker is a small standalone pilot job that waits for instructions from the TaskVine scheduler and operates duly. Once tasks are completed, workers communicate the results back to the scheduler, which forwards them back to the application level. 
The TaskVine scheduler does not delegate the local resource management to individual workers: each task comes with a specific amount of resource allocation, and each worker is directed by the TaskVine scheduler on how to utilize any local resource type (CPU, memory, SSD, GPU). The pool of resources is maintained by the TaskVine factory, a daemon-like process that monitors the current resource pool and adjusts it based on a given resource policy and the current load of the cluster.

Given this software stack, it is then straightforward for a user to scale up a throughput-oriented inference application.  
To satisfy Condition \#1, a user first defines an arbitrary computation involving LLM inferences in a Python function. This function then flows through Parsl and the TaskVine scheduler to a TaskVine worker as a task to be executed. 
Each worker is allocated with a small number of GPUs such that a given task can run comfortably. Once the task completes, inference results are sent from the worker back to the application as described above. 
Condition \#2 is inherently satisfied by the design of the software stack: the scheduler has a queue of ready tasks, and its main job is to occupy connected workers with tasks at any given time. Therefore, the application will make progress as long as there are workers connected to the scheduler. Condition \#3 is reasonably assumed in the cluster as discussed.

Finally, this software stack provides a seamless integration with opportunistic resources. The scheduler on the manager node directs all workers on what to do and thus keeps a globally consistent view of the application. This means that workers can leave and join the pool freely as tracked by the TaskVine scheduler and adjusted by the TaskVine factory, and any evicted task is detected, retrieved, and re-inserted into the queue of ready tasks by the scheduler. This software stack thus addresses Challenge \#1 by design. Challenge \#2 is satisfied by the maturity of Parsl and TaskVine as individual pieces of open-source software: each has its own suite of performance observability tools to help users visualize the throughput rate and progress of an application. Note that Challenges \#3-6 also need to be addressed to achieve a performant scaling of inference applications on opportunistic resources.

\subsection{Pervasive Context Management}
\label{subsesc:pervasive}
\begin{figure}[t]
    \includegraphics[width=\columnwidth]{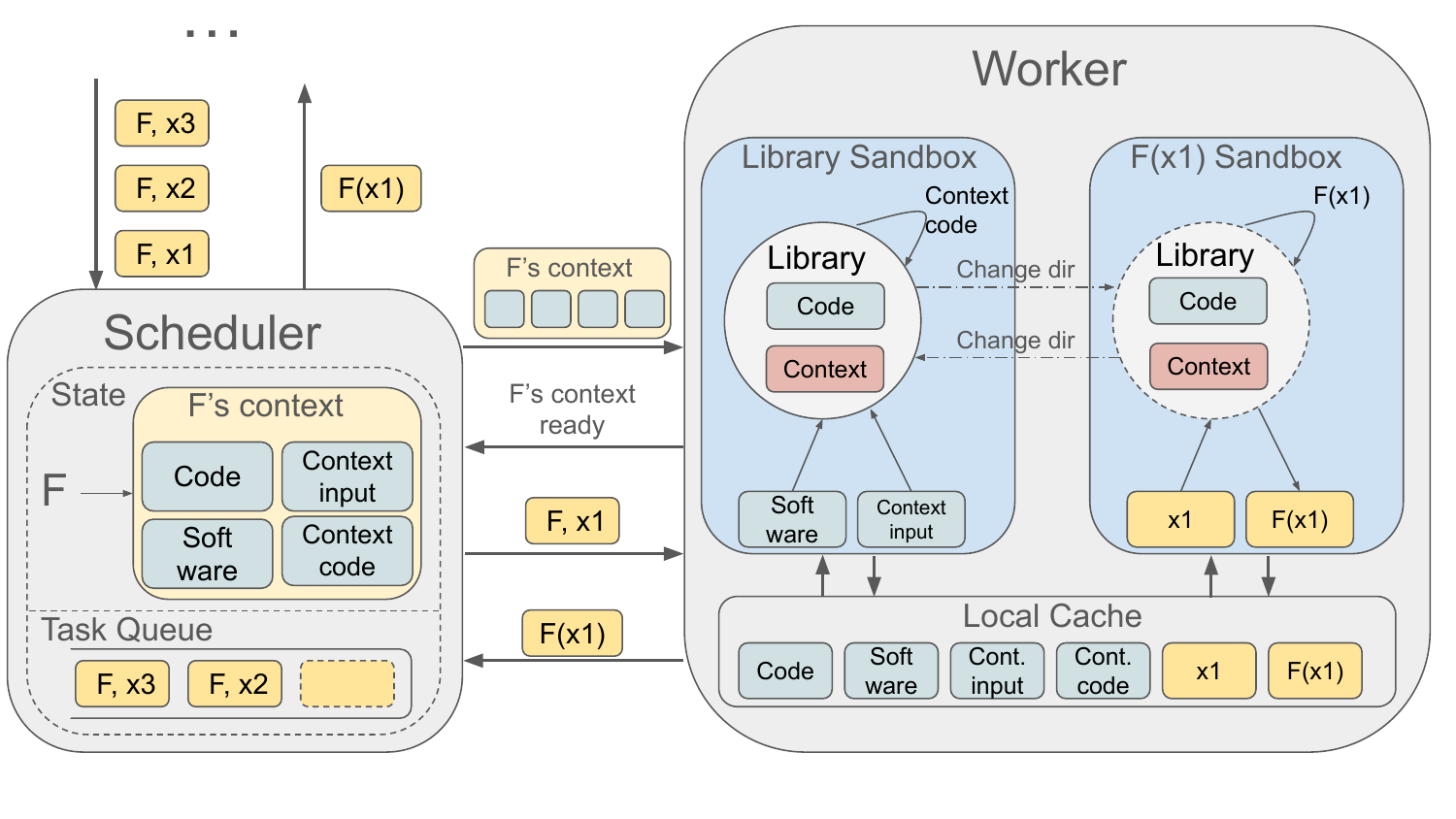}
    \caption{Context Reuse with Pervasive Context Management}
    \justifying
    \label{fig:context-mgmt}
    \vspace{-.2in}
\end{figure}
The second component of our solution is the pervasive context management technique that allows efficient reuse of a common computational context between tasks. 
We first point out the 
three following observations: (1) each task almost always needs to set up some computational context before any work is done, and thus carries some non-trivial initialization overheads (e.g., importing relevant libraries, constructing a local database, moving an LLM from SSDs or memory to GPU), (2) a large portion of this context is shareable and reusable between tasks as they usually follow the same code path and only diverge at the execution of individual inputs, and (3) since tasks are designed to be independent (see Section \ref{sec:background}), each creates its own context for execution and tears the context down in the clean up phase, which forces each task to pay the same cost of context initialization and misses out on the opportunity of sharing and reusing such context. 
The core idea of pervasive context management is then to capitalize on this opportunity: a reusable context is extracted from each function, and subsequent invocations of that function can utilize this context to speed up its computations. Note that a context, in this sense, is an arbitrary computational state, which can be hosted on any worker in the pool of resources and can materialize in any format (disk, memory, GPU). This subsection focuses on the latter that concerns reusing an arbitrary context on a given worker, and delays the discussion on context distribution to Subsection \ref{subsec:support}.

\begin{figure}[t]
\begin{minted}[fontsize=\small, linenos]{python}
from parsl import python_app
def load_model(model_path):
    ...
    model = AutoModel.from_pretrained(model_path).to('gpu')
    return {'model': model}

@python_app
def infer_model(inputs, parsl_spec):
    from parsl import load_variable_from_serverless
    model = load_variable_from_serverless('model')
    outputs = [model.generate(input) for input in inputs]
    return outputs

model_path = ...
parsl_spec = {'context': [load_model, [model_path], {}]}
inputs = ...
results = infer_model(inputs, parsl_spec).result()
\end{minted}
\vspace{-.1in}
\caption{Code Example of an LLM Inference Application}
\label{fig:sample-code}
\RaggedRight
\vspace{-.2in}
\end{figure}

Figure \ref{fig:context-mgmt} gives a general example of how context reuse works in the Parsl-TaskVine stack. An application (not shown) starts up and invokes function F three times with arguments x1, x2, and x3, respectively. Parsl (not shown)  sees that these three functions don't have any dependency between them and converts them into ready tasks to be sent to the TaskVine scheduler. The scheduler examines F and discovers its context recipe, including F's code, software dependencies, context code, and context inputs, (we discuss the context discoverability mechanism later in Subsection \ref{subsec:support}) and sends this context recipe to be hosted on a given worker as part of F(x1) execution. The worker, upon receiving the context recipe, stores all of its components in a local cache and fork-execs a special process called library. The library process is responsible for materializing and hosting F's context from its recipe and will cooperate with the worker to execute subsequent invocations of F. 
Upon the stage-in of F's context into its sandbox, the library registers F's code, executes the context code, stores the resulting context as an internal state in its process, and lets the worker know it's ready for invocations of F. The scheduler, upon receiving 
this ack from the worker, sends the first invocation request of (F, x1). 
The worker stores x1 in its cache, creates a sandbox for the invocation, and pings the library. The library then changes its working directory to F(x1)'s sandbox and executes the invocation directly in its address space, which already contains F's context, before returning to its sandbox. The result of the invocation is then returned to scheduler, which marks the completion of F(x1) and forwards the result to the application. Executions of (F, x2) and (F, x3) then reuse F's context via the library and follow the same path (F, x1) took.

The pervasive context management technique addresses Challenges \#3, 5, and 6 as follows. For Challenge \#3, the hosting of F's context acts as a cache of the initialization overhead: instead of unnecessarily repeating the same initialization computation, the cost of putting up a common context is paid once as part of the first function invocation and amortized by subsequent invocations of the same function. Challenge \#5 is addressed by the localization of individual tasks' I/O: TaskVine stages in all inputs a function needs and stages out all outputs a function produces via workers' local caches, thus removing the reliance on a shared filesystem. Finally, Challenge \#6 is addressed by the fact that a cost of creating a common context is paid only once, no matter the batch size. Specifically, the overhead is constant per worker with pervasive context management, and a given application's runtime now depends only on the net throughput each worker can produce.

Figure \ref{fig:sample-code} shows a 
code example of how an inference application can utilize this technique in the Parsl-TaskVine stack. Lines 2-5 define a \verb|load_model| function that creates an LLM context by loading its parameters from disk to GPU and returns this context via a dictionary. This dictionary informs the library of the relevant context to later be exposed to the actual invocation. Lines 7-12 define the actual computation via the \verb|infer_model| function that loads the model from the context held by the library, executes the inferences, and returns the results. Lines 14-17 connect the missing pieces of the example where the context computation is defined via the \verb|parsl_spec| variable, and \verb|infer_model| brings this context reference along with its inputs to the scheduler for execution.

\begin{table}[t]
    \centering
    \begin{tabular}{|c|c|c|}
        \hline
         Device Name &  Release Year & Count \\
        \hline
               NVIDIA Quadro RTX 6000 & 2018 & 106\\
        \hline
        NVIDIA A10 & 2021 & 78 \\
        \hline
        NVIDIA TITAN X (Pascal) & 2016 & 69 \\
        \hline
        NVIDIA GeForce GTX 1080 Ti & 2017 & 63 \\
        \hline
        NVIDIA RTX 6000 Ada Generation & 2022 & 36 \\
        \hline
        NVIDIA GeForce GTX TITAN X & 2015 & 34 \\
        \hline
        NVIDIA A40 & 2020 & 26 \\
        \hline
        NVIDIA H100 80GB HBM3 & 2023 & 15 \\
        \hline
    \end{tabular}
    \caption{8 Major GPU Models in the Local Cluster}
    \label{tab:gpus}
    \vspace{-.4in}
\end{table}

\subsection{Supporting Mechanisms and Policies}
\label{subsec:support}
\subsubsection{Mechanisms}
We now describe the supporting mechanisms that enables pervasive context management, focusing on the discoverability and distribution of a function's context between connected workers. Note that the context retention mechanism is described in detail in Subsection \ref{subsesc:pervasive}.

\begin{figure*}[t]

\includegraphics[width=18cm, height=6cm]{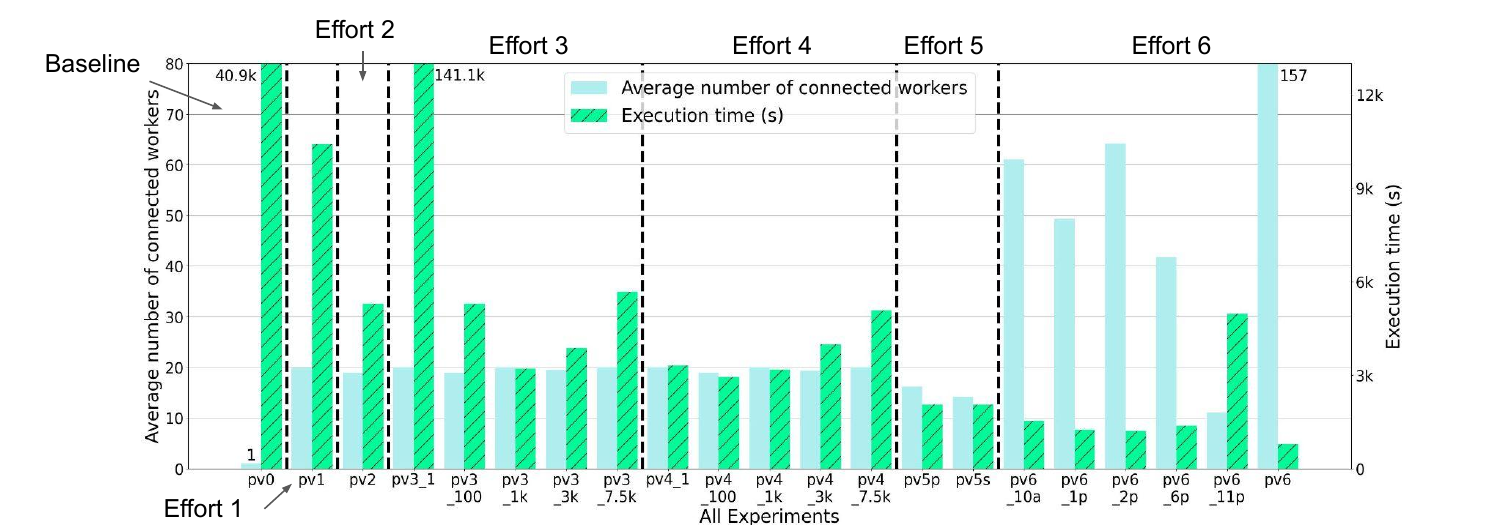}
\vspace{-.3in}
\caption{Average Number of Connected Workers and Execution Time of All Experiments} 
\it{All 21 experiments are grouped in 6 incremental efforts from left to right. Later efforts generate better results, subject to the number of workers.}
\label{plt:all}
\vspace{-.1in}
\end{figure*}

A computational  context consists of 4 elements: the function's code, its software dependencies, the context code, and the context inputs. To discover a function's context, we first use the Poncho toolkit\cite{sly2022poncho} to pack the function's software dependencies into a portable format such that the scheduler can send this package to any worker. The function's code is serialized via the cloudpickle\cite{cloudpickle} module and sent to the worker, which is then deserialized by the library back into a Python code object. Finally, we rely on the user to manually create the context code and its inputs and bind it to the function via Parsl primitives, and these elements also undergo the same serialization and deserialization process.

To distribute the context quickly between workers, we use the built-in peer transfer feature from TaskVine. This feature allows workers to communicate and send arbitrary data to each other under the direction of the scheduler, and each worker is capped at N transfers at a given moment. The context distribution then takes the shape of a spanning tree: the scheduler first sends the context to an arbitrary worker, and this worker sends the context to N other workers, and so on until the context is fully distributed and hosted across all workers in the system.

\subsubsection{Policies}
In the Parsl-TaskVine software stack, a TaskVine worker is the base unit of resource acquisition: each worker holds a certain amount of resources, and resources join and leave the application's pool via the instantiation and eviction of these workers. The rate at which resources join and leave thus depends on how resource requirements of each worker are designed. On one hand, a user wanting a higher rate of resource acquisition would submit a smaller number of larger workers as each worker instantiation instantly grabs a large chunk of available resources. On the other hand, a user who wishes to alleviate the effect of worker eviction would prefer submitting many smaller workers as resources lost due to eviction are more fine-grained as opposed to losing a large chunk of resources per worker.

When inference applications are run on opportunistic resources, eviction becomes more common and thus is a larger concern. Our policy consequently is the latter approach where we design the resource requirements of each worker to be as small as possible to better conserve the existing resource allocation. Thus, each worker is submitted independently in a different batch job with a minimal amount of resource requirements, and once instantiated, runs at most 1 task at any given time. This 1-to-1 ratio between tasks and workers also helps addressing Challenge \#4 as workers that land on compute nodes with better hardware (e.g., GPU) run tasks faster and thus run more tasks compared to workers with slower GPUs. The former approach, while having a better rate of resource acquisition, risks binding too many tasks to the slower workers and thus exhibits the undesirable straggling effect.

\section{Evaluation}
\label{sec:eval}

This section demonstrates the effectiveness of pervasive context management on executing a given throughput-oriented LLM inference application on opportunistic resources. We curate this section as a series of incremental scaling efforts and 
interweave in-depth discussions about our reasoning, results, explanations, and potential pitfalls. We first give a thorough description of the inference application we scaled along with the general settings that apply to all experiments,
and finally present our scaling efforts.

\subsection{Optimal Prompt Search in Fact Verification}
Fact verification is an active area of research given the lightning rise of online mis- and dis-information\cite{zhang2023towards, gunjal2024molecular}.
Our application, Prompt for Fact (PfF), uses a given LLM as a fact verifier to check the correctness of an arbitrary claim. Since there are many LLMs to try as a fact verifier and even more abundant prompting strategies, PfF seeks to find an optimal pair of (LLM, prompt template) that yields the highest  accuracy in a particular fact verification dataset. 
We implemented an MVP of this application that takes an LLM that satisfies Condition \#1 and a prompt template and returns the aggregated accuracy. Extending this MVP to many LLMs and prompt templates is straightforward as each combination is independent  from one another and thus fully parallelizable. 

Specifically, we use the training data from FEVER\cite{Thorne18Fever} as our dataset containing 145,449 claims, each of which is labeled with either \verb|SUPPORTED|, \verb|REFUTED|, or \verb|NOT ENOUGH INFO|. Each claim contains a statement about a given subject and a list of references to relevant Wikipedia pages.
Per the LLM, we use the recently released SmolLM2 model with 1.7 billion parameters\cite{allal2025smollm2}.
Our MVP takes the LLM and a prompt template, runs a full inference sweep across the dataset, and returns the fact verification accuracy. Note that the MVP also satisfies Condition \#2 as we only care about the aggregated  accuracy over the whole dataset.

Additionally, to reflect the progression of our understanding of pervasive context management, we differentiate two types of context: partial context containing only software dependencies and model parameters, and pervasive context that includes software dependencies, inference code, context code, and context inputs. 

\vspace{-.1in}

\subsection{General Experiment Settings}
Our local cluster runs two resource managers: it uses the Altair Grid Engine (AGE)\cite{age} as the main static batch manager, and runs HTCondor\cite{thain2005distributed} as a resource backfiller on AGE's unallocated machines. There are 567 GPUs in total in the cluster with 18 different models. Table \ref{tab:gpus} shows 8 major GPU models that account for 75\% of all GPUs in the cluster. Each model is annotated with its release year and the number of GPUs in that model, showing the evolution of our local GPU cluster over time and the fulfillment of Condition \#3. We run all experiments through the HTCondor resource manager. Our cluster provides access to data via the Panasas ActiveStor 16\cite{shaffer2017taming,panasas} shared filesystem with 77 nodes and supports up to 84 Gbs/s read bandwidth and 94k read IOPS.

\begin{figure}[t]
\begin{minipage}[t]{0.475\columnwidth}
  \includegraphics[scale=0.28]{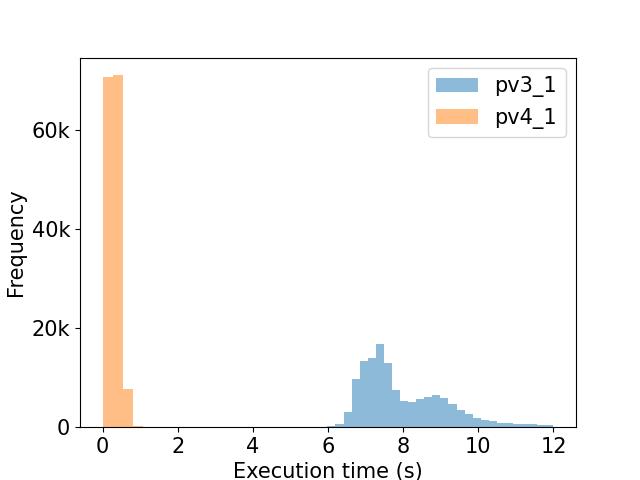}
\end{minipage}
\begin{minipage}[t]{0.475\columnwidth}
  \includegraphics[scale=0.28]{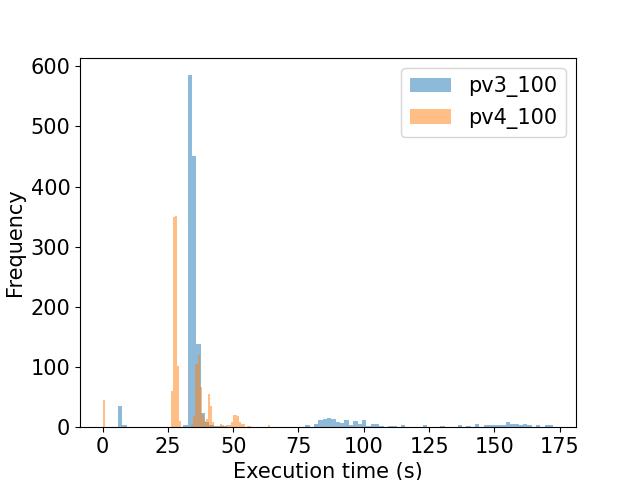}
\end{minipage}
\caption{Effect of Pervasive Context on Task Exec. Time}
\textit{Pervasive context (pv4\_[1, 100]) results in faster and more predictable execution times of tasks, compared to partial context (pv3\_[1, 100]).}
\label{plt:hist}
\end{figure}

We configure parameters of our Parsl-TaskVine software stack as follows. We enable the peer transfer feature that allows workers to communicate and send arbitrary data between each other. Each task's resource allocation includes 2 cores, 10 GBs of memory, 20 GB of disk, and 1 GPU, providing a comfortable amount of resources for a smooth inference execution. Each TaskVine worker has 2 cores, 10 GBs of memory, 70 GBs of disk, and 1 GPU, thus providing the worker with just enough resources to run tasks in a 1-to-1 manner to preserve claimed opportunistic resources and plenty of disk space for local caching.

\begin{table}
    \centering
    \begin{tabular}{|c|c|c|c|c|}
        \hline
        Exp. ID  & Mean &  Std. Dev. & Min & Max \\
        \hline
        pv3\_1 &  15.10 & 27.26 & 5.55 & 390.03 \\
        \hline
\textbf{pv4\_1} &  \textbf{0.32} & \textbf{0.13} & \textbf{0.0008} & \textbf{15.25} \\
        \hline
        pv3\_100  & 46.78 & 32.88 & 5.93 & 195.89 \\
        \hline
        \textbf{pv4\_100} & \textbf{31.91} & \textbf{9.3} & \textbf{0.0008} & \textbf{79.05} \\
        \hline

    \end{tabular}
    \caption{Statistics of Tasks' Execution Time in 4 Experiments}
    \vspace{-.15in}
    \it{Pervasive Context (in bold) greatly reduces  statistics of tasks' execution time in experiments with smaller batch sizes.}
    \label{tab:task_stats}
    \vspace{-.2in}
\end{table}

Almost all experiments start with the same resource pool configuration consisting of 20 GPUs, where half are NVIDIA A10 and the other half are NVIDIA TITAN X (Pascal). This approach allows us to not only establish consistency and stability to our measurements and results but also mimic the heterogeneity of the actual GPU cluster. An experiment starts when 95\% of all GPUs join the pool.
These constraints are removed at the end which allow the application to have access to up to
186 opportunistic GPUs.

Finally, we present the parameters of our application. Since the original FEVER dataset contains references of relevant Wikipedia pages in each claim, a data joining is required to resolve these references to the exact source text. We preprocessed this joining and store it in a local database, and all experiments pull fully resolved claims from this database to speed up the process. We introduce a small number of empty claims as the control group and bring the total number of inferences to 150k. Each inference task carries a batch size of 100 claims (thus 100 inferences) by default. This batch size is arbitrarily chosen and we provide our analysis on optimal batch sizing later. Storage-wise, the LLM takes up 3.7 GBs of disk and around 7.4 GBs of memory when fully loaded. The application's software dependencies are managed in a Conda\cite{conda} environment, containing 308 packages and totalling 10.5 GBs of disk. The Poncho package of this environment brings the disk size down to 3.7 GBs.  
Note that these settings apply to all experiments unless explicitly
noted otherwise.

\subsection{Scaling Efforts}
Figure \ref{plt:all} shows the average number of connected workers and the execution time of all experiments. Each experiment has a specific ID, and our incremental efforts are shown from left to right. We report the average number of connected workers as worker evictions and joining are common in HTCondor and happen in almost all experiments (both involuntarily and intentionally). A closer look into worker evictions and joining is provided in the later efforts. Note that from this point, we implicitly refer to Figure \ref{plt:all} whenever we discuss any set of experiments.

\begin{figure}[t]

\includegraphics[width=8.3cm, height=5.4cm]{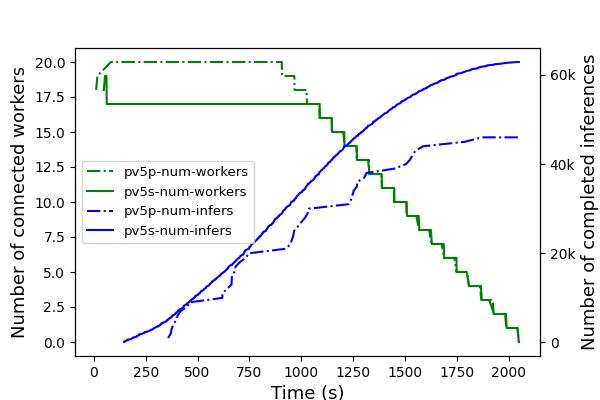}
\caption{Effect of Pervasive Context on Throughput}
\textit{Use of pervasive context (pv5s) results in
36.7\% more work done than partial context (pv5p) in a busy cluster with diminishing availability.}
\label{plt:done_infer}
\vspace{-.1in}
\end{figure}

\begin{figure*}[t]
\begin{subfigure}[t]{0.325\textwidth}
\includegraphics[scale=0.27]{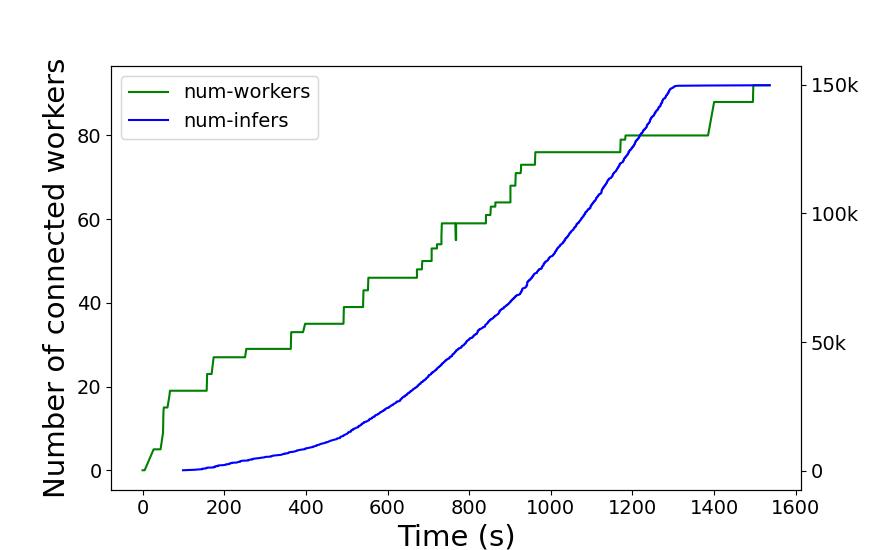} 
  \captionsetup{justification=centering}
 \end{subfigure}
 \begin{subfigure}[t]{0.325\textwidth} 
 \includegraphics[scale=0.27]{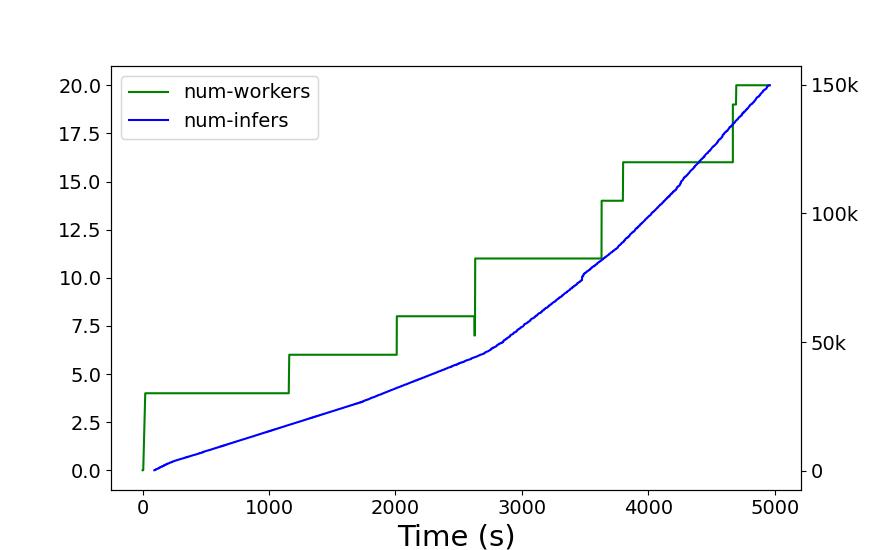} 
 \captionsetup{justification=centering}
 \end{subfigure}
 \begin{subfigure}[t]{0.325\textwidth}
 \includegraphics[scale=0.27]{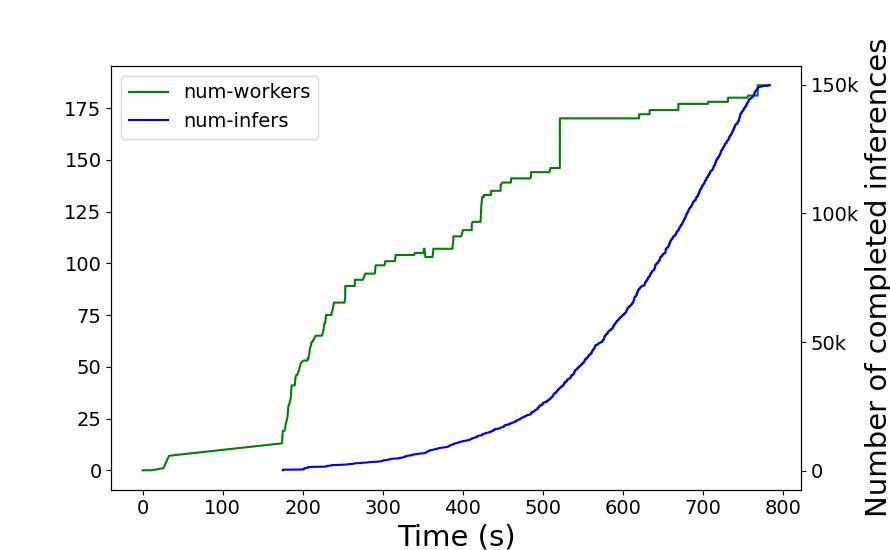}
 \captionsetup{justification=centering}
 \end{subfigure}
\caption{Application Resilience Against Dynamic Opportunistic Resources for pv6\_10a (left), pv6\_11p (middle), and pv6 (right)}

\textit{ Note that plots share the right y axis but have their own scales of the x axis and the left y axis. Application's inference progress seamlessly adapts to the availability of opportunistic resources (represented via connected workers) in all cases.}
\label{plt:unres}
\vspace{-.15in}
\end{figure*}

\textbf{Baseline: 1 GPU [pv0].} The pv0 experiment sweeps

of the fact verification dataset on 1 worker with a dedicated NVIDIA A10 GPU as our baseline, which takes 40.9k seconds from start to finish.

\textbf{Effort 1: Naive Scaling [pv1].}
The pv1 experiment shows the results when we scale the application to 20 GPUs using a naive implementation of the Parsl-TaskVine stack. In this implementation, we minimize the amount of code changes compared to the baseline, and only divide the dataset into 1,500 tasks, each with 100 inferences, and run these tasks in parallel on 20 workers. This effort thus produces a disappointing speedup of 3.9 (from 40.9k to 10.4k seconds) on 20 GPUs due to several reasons. 
Since all inference tasks run independently and register no reusable context, the peer transfer feature is effectively disabled as there's no registered data to be transferred between workers. All software dependencies are pulled from the shared filesystem so the I/O performance depends on its current load. Additionally, each task downloads its own copy of the model from the Internet to its local execution sandbox instead of reusing a local cache of model parameters as sandboxes are created upon execution and destroyed upon cleanup.

\textbf{Effort 2: Software and LLM as Partial Context [pv2].}
Experiment pv2 shows the next effort that registers software dependencies and model parameters as partial context to the inference tasks, bringing the speedup to the factor of 7.7 (from 40.9k to 5.3k seconds). Note that this effort uses an inference batch size of 100, and our next effort searches for an optimal batch size.

\textbf{Effort 3: Partial Context Management with Batch Size Tuning [pv3*].}
Experiments pv3\_[1, 100, 1k, 3k, 7.5k] represent our search for an optimal batch size with partial context management, where each experiment's batch size is the suffix of its ID.
Figure \ref{plt:all} shows a parabolic pattern of execution times as both ends show much higher values than pv3\_1k. 
This pattern is due to two competing factors: resource heterogeneity and initialization overheads. On one hand, a larger batch size better amortizes the  overhead of creating a model state and moving it to a GPU. Over-batching inferences in a task however risks running many inferences on slower GPUs due to resource heterogeneity, which increases the execution time of the application with pv3\_7.5k being the most extreme case of this effect. 
Specifically, since we have 150k inferences and 20 GPUs, a batch size of 7.5k divides all inferences equally into 20 batches where each batch is run in a GPU on a worker. 
pv3\_7.5k's execution time thus equals to that of the slowest GPU, no matter how fast the other 19 GPUs are. pv3\_3k only alleviates this effect with a smaller batch size which results in a slightly better execution time. On the other hand, a smaller batch size helps spread more inferences to faster GPUs but risks paying a costly penalty of many more initialization overheads. pv3\_1, the other end of the spectrum demonstrating this detrimental effect, runs only 1 inference per model creation and loading which substantially increases the execution time to 141.1k seconds. 
With partial context management, pv3\_1k empirically becomes the best run with a batch size of 1,000 that is both small enough to distribute more computations to fast GPUs and large enough to amortize the overhead cost. We also observe two competing factors, resource heterogeneity and initialization overheads, as important considerations for next efforts. 

\textbf{Effort 4: Pervasive Context Management with Batch Size Tuning [pv4*].}
Since resource heterogeneity is an intrinsic feature of opportunistic resources, effort 4 instead focuses on eliminating the initialization overheads with pervasive context management.
Each worker now only pays a one-time cost of model creation and loading that allows overhead amortization regardless of batch sizes.
Experiments pv4\_[1, 100, 1k, 3k, 7.5k] in Figure \ref{plt:all} show the execution times of the application with pervasive context management on respective batch sizes. While the execution times of pv4\_[1k, 3k, 7.5k] don't considerably change as their batch sizes shield them from the overheads' penalty (pv4\_7.5k is slightly better than pv3\_7.5k as the latter has 1 worker eviction), the performance of pv4\_[1, 100] are much better than their pv3 counterparts by 97.8\% and 44.5\%,  respectively, and drives down the fastest execution time to 2.9k seconds with a speedup factor of 13.9. 
We point out two observations: 1) the optimal batch size now shifts from 1k to 100, showing that an even finer-grained batch size can benefit from pervasive context management, and 2) the performance cost of choosing the wrong batch size decreases significantly as any batch size in the range of 1 to 1,000 now results in an execution runtime increase of at most 12.3\% instead of 4306\%.
Note that the ideal speedup factor of 20 is impossible as the pool of resources is heterogeneous.
Moreover, pervasive context management enables much lower and more stable execution times of individual tasks in runs with smaller batch sizes.
Figure \ref{plt:hist} shows the histograms of execution time of two pairs of runs, pv[3, 4]\_1 (left) and pv[3, 4]\_100 (right). On the left histogram, two runs both have 150k tasks, each with 1 inference, but pv3\_1 employs partial context management while pv4\_1 follows pervasive context management. This difference results in the stark contrast of tasks' execution time, as most tasks in pv4\_1 reuse existing contexts on workers more effectively and cluster around the (0, 1) second range, while tasks in pv3\_1 must pay the cost of model loading and mostly spread from 6 to 12 seconds. A similar pattern shows in the right histogram with the pv[3, 4]\_100 pair as pv3\_100 has a much wider spread of execution time compared to pv4\_100. Note that we trim some values in histograms for better visualization. 
Table \ref{tab:task_stats} instead shows the full statistics of tasks' execution time of 4 runs, demonstrating the superiority of pervasive context management as it has lower mean, standard deviation, min, and max execution times in both pairs.

\textbf{Effort 5: Pervasive Context Management In a Busy  Cluster  [pv5*].}
We describe the last set of experiment on the 20-GPU setup before running the application without any restriction on the local cluster. This last set contains pv5p and pv5s which follow the partial and pervasive context management approaches and have the empirically optimal batch sizes of 1k and 100 respectively, and simulates an eviction scenario where a GPU cluster suddenly becomes busy and reclaims opportunistic resources gradually. Specifically, both experiments run without any artificial interruption for first 15 minutes, and then we drain all GPUs from the resource pool at the rate of 1 GPU/minute, prioritizing all NVIDIA A10s before NVIDIA Titan X Pascals.
Figure \ref{plt:done_infer} shows the number of connected workers and the amount of completed inferences over time under the two context management policies.  
Even though pv5s involuntarily loses 2 workers from the beginning, its throughput rate is still consistently higher than that of pv5p at any given time. We thus see another benefit of pervasive context management besides the constant cost of overhead per worker: while both experiments eventually have 20 workers and thus 20 tasks evicted, the amount of inferences evicted in pv5s is only 20 * 100 = 2k while that of pv5p is 20* 1000 = 20k. This results in a striking difference of 16.9k completed inferences at the end of the experiments, and thus demonstrates the effectiveness of the pervasive context management technique.

\textbf{Effort 6: Unrestricted Scaling [pv6*].}
Experiments pv6\_[10a, 1p, 2p, 6p, 11p] represent the application running with pervasive context management and a batch size of 100 in a given day on the local cluster without any restriction, with each experiment's suffix denoting the time of day the application starts (from 10am to 11pm). We can see that the amount of claimed opportunistic resources on average fluctuates by a large margin in the day, going as high as 64 to as low as 11 GPUs, with later experiments showing fewer opportunistic GPUs as users tend to run more jobs overnight. 
Nevertheless, the application's execution time adapts proportionally to the instability and unpredictability of the opportunistic GPUs, with pv6\_2p's execution time going as low as 1,211 seconds. We also show that pv6, an experiment with the same settings but run on a different day when the cluster is less busy, is able to claim an average of 157 GPUs during its shortest execution time of 783 seconds. Figure \ref{plt:unres} shows more details of 3 experiments, pv6\_10a, pv6\_11p, and pv6, plotting the number of connected workers and completed inferences over time. 
We can observe the different rates of resource acquisition during the application's execution, showing the uncertainty of the opportunistic resources. However, the throughput rate of the application doesn't lag too far behind the amount of acquired resources and adapts seamlessly to the availability of opportunistic resources in all cases. This final effort thus demonstrates the success of pervasive context management in scaling up a throughput-oriented  inference application on a heterogeneous opportunistic GPU cluster.

\vspace{-.1in}

\section{Related Works}
\label{sec:related}

\textbf{LLM Inference Optimization.}
Many works optimize the inference process by outputting several tokens in one forward pass based on the speculative decoding scheme\cite{leviathan2023fast,spector2023accelerating, chen2024cascade, svirschevski2024specexec}. This scheme assumes that an LLM generating tokens sequentially takes too much time and resources, especially with easy-to-predict tokens. To speed this up, a smaller LLM is used to predict the next K tokens in advance, and the original LLM can make one forward pass that accepts tokens it agrees with and rejects others instead of making K forward passes.
Other works focus on KV cache and memory management on both a single GPU and a pool of GPUs. Kwon et. al.\cite{kwon2023efficient} introduce a virtual paging mechanism that divides the dynamically-sized KV cache into blocks to remove GPU memory fragmentation, while Lin et. al.
\cite{lin2024infinite} distribute the KV cache and the attention computation to many GPUs.
Cloud deployment of inference serving is also an active area of research. Fu et. al.\cite{fu2024serverlessllm} use local storage of individual instances to cache and distribute model checkpoints among each other. Our work extends the usage of local storage to memory and GPUs to hold and distribute the computational context. Mao et. al. \cite{mao2025skyserve} mix on-demand (equivalent to static resources in clusters) with spot instances (generally equivalent to opportunistic resources) and over-provision spot instances to serve latency-sensitive LLM inferences. Our work runs LLM inferences only on opportunistic resources instead. Miao et. al.\cite{miao2024spotserve} also exclusively serve LLM inferences on spot instances, but rely on a grace period from 30 seconds to 2 minutes to send the state of an ongoing request to other instances, while opportunistic resources in our work evict workers immediately upon reclamation.

\textbf{Workflow Systems.}
Workflow systems evolve from the traditional resource managers and allow applications to express complex relationships between tasks via a directed acyclic graph (DAG) instead of a bag of tasks\cite{deelman2015pegasus, turilli2019middleware, zheng2017deploying}. These systems typically focus on applications' reliability, performance, and portability via novel architectural designs and runtime optimizations, but require users to describe the computational needs in detail via complicated and non-uniform abstractions. More modern workflow systems\cite{babuji2019parsl, rocklin2015dask, moritz2018ray} tackle this usability problem by providing Pythonic abstractions that enable users to wrap their computational needs neatly into Python functions and translating these functions into tasks deployable to remote nodes. Our Parsl-TaskVine integration follows this movement and allows users to easily describe their computations in Python without losing performance, reliability, or portability. 
The Parsl-TaskVine stack extends this movement one step further with the support of computational context sharing between tasks on contrary to the traditional view of complete inter-task independence. Ray\cite{moritz2018ray} is a Python workflow system that also relaxes the independence abstraction and offers a similar sharing abstraction via Ray Actors. 
Specifically, Ray Actors allow users to define an object with associated methods such that users can deploy these objects on compute nodes and remotely invoke their methods in a similar fashion to Java RMI\cite{rmi}. Thus, users of the Parsl-TaskVine stack defining a computational context can equivalently do so by putting the context code in the object's class constructor, which both holds the context on compute nodes as long as the object is alive and provides tasks access to this context via the object's Python reference. The major difference between the two approaches lies in their abstraction as Parsl-TaskVine supports the task abstraction while Ray supports the object abstraction. In terms of scheduling, Parsl-TaskVine users only have to submit tasks and the system automatically maps tasks to available contexts in the cluster. 
On the other hand, Ray users must explicitly choose an object for every remote task execution. This also has an implication in reliability. When a worker holding a context is evicted, Parsl-TaskVine seamlessly requeues the task and moves it to another node and shields users from this failure, while Ray users must manually deal with an actor eviction before sending tasks for remote execution.

\section{Conclusion}
The LLM technology has been improving at an astonishing rate over the years and promises an unprecedented leap in human productivity. The current infrastructure however is ill-equipped to deal with the massive spike in interests and computational demands from LLM scientists and practitioners and requires new approaches to efficient resource management. This paper analyzes the resource requirements of traditional LLM training and inference serving applications, points out a class of LLM applications that relaxes the latency and synchronization requirements, and
introduces the pervasive context management technique that allows this class of inference applications to seamlessly and performantly scale on heterogeneous opportunistic GPU clusters. Specifically, it points out three conditions that enable a throughput-oriented execution of inference applications and six challenges in effective utilization of opportunistic resources, followed by in-depth descriptions of the Parsl-TaskVine stack and pervasive context management as the solution, and a thorough demonstration of our scaling efforts, resulting in a 98.1\% reduction in execution time of a given LLM inference application.

\newpage

\bibliographystyle{ACM-Reference-Format}
\bibliography{main}

\end{document}